\documentclass[twocolumn,showpacs,superscriptaddress,prb]{revtex4-1}
\usepackage{amsfonts}
\usepackage{amsmath}
\usepackage{amssymb}
\usepackage{graphicx}

\begin{document}

\title[Photon-assisted scattering]{Photon-assisted scattering and magnetoconductivity oscillations in
a strongly correlated 2D electron system formed on the surface of liquid helium}
\author{Yu.P. Monarkha}
\affiliation{Institute for Low Temperature Physics and Engineering, 47 Lenin Avenue, 61103
Kharkov, Ukraine}

\begin{abstract}
The influence of strong internal forces on photon-assisted scattering and on
the displacement mechanism of magnetoconductivity oscillations in a
two-dimensional (2D) electron gas is theoretically studied. The theory is
applied to the highly correlated system of surface electrons on liquid
helium under conditions that the microwave frequency is substantially
different from inter-subband resonance frequencies. A strong dependence of
the amplitude of magnetoconductivity oscillations on the electron density is
established. The possibility of experimental observation of such
oscillations caused by photon-assisted scattering is discussed.
\end{abstract}




\maketitle

\section{Introduction}

Radiation-induced magnetoresistivity oscillations and zero resistance states
(ZRS) of a 2D electron gas subjected to a perpendicular magnetic field
were discovered in 2001-2003 using
high quality GaAs/AlGaAs heterostructures~\cite{ZudSim-2001,ManSme-2002,ZudDu-2003}.
Since then, a large number of
theoretical mechanisms have been proposed to explain these magneto-oscillations
(MO)~\cite{DurSac-2003,RyzSur-2003,Shi-2003,KouRai-2003,RyzChaSur-2004,DmiVav-2005,InaPla-2007,Mik-2011}.
The ZRS appeared at high radiation power as a result of
evolution of resistivity minima can be
caused by a negative conductivity effect ($\sigma _{xx}<0$)~\cite{AndAle-2003}, whose
microscopic origin  is quite controversial as well as the origin of MO.

In experiments~\cite{ZudSim-2001,ManSme-2002,ZudDu-2003}, the microwave (MW) frequency
was quite arbitrary: $\omega >\omega _{c}$, here $\omega _{c}$ is the
cyclotron frequency. The period of
MO observed is controlled by the ratio $\omega /\omega _{c}$. Similar $1/B$-periodic
oscillations of magnetoconductivity $\sigma _{xx}$ and ZRS were
discovered in a nondegenerate 2D electron system formed on the free
surface of liquid helium when the MW frequency was tuned to the
inter-subband excitation frequency~\cite{KonKon-2009,KonKon-2010}: $\omega =\omega _{2,1}\equiv
\left( \Delta _{2}-\Delta _{1}\right) /\hbar $ (here $\Delta _{l}$ is the
energy spectrum of surface subbands, $l=1,2,...$). These oscillations were
theoretically explained~\cite{Mon-2011} by a nonequilibrium population of the excited subband
and by peculiarities of quasielastic intersubband scattering in the presence of
the magnetic field $B$. Predictions of this intersubband displacement
model concerning the influence of internal Coulomb forces on positions of $%
\sigma _{xx}$ extrema~\cite{Mon-2012} were recently supported by experimental
observations~\cite{KonMonKon-2013}.

For mechanisms of MO proposed~\cite{DurSac-2003,RyzSur-2003,DmiVav-2005},
electron gas degeneracy
is not a crucial point. Therefore, they can be applied also
to a nondegenerate 2D electron system like surface
electrons (SEs) on liquid helium. Electrons bound to the free surface of
liquid helium represent a remarkable model 2D system which is quite simple
and clean. SEs are scattered quasi-elastically by capillary wave quanta
(ripplons) and by vapor atoms. Their interaction parameters are well
established. Experiments on SEs~\cite{KonKon-2009,KonKon-2010}
employed approximately the same
MW frequencies and power as those used for the 2D electron
gas in GaAs/AlGaAs~\cite{ZudSim-2001,ManSme-2002,ZudDu-2003}.
Therefore, there is an important question: why
theoretical mechanisms of MO and negative conductivity effects proposed for
semiconductor electrons do not display themselves in experiments with SEs on
liquid helium? The answer to this question could shed light also on the
situation in semiconductor systems.

The most frequently discussed mechanism of MO and negative
conductivity effects called the displacement mechanism was proposed
already in 1969 by Ryzhii~\cite{Ryz-1969}. In this model a quasielastic scattering
event of an electron caused by an impurity potential can be accompanied by
absorption of a photon, which leads to indirect inter-Landau-level scattering
($n\rightarrow n^{\prime }$). The energy conservation of such a
photon-assisted scattering event
\begin{equation}
\hbar \omega _{c}\left( n-n^{\prime }\right) +\hbar \omega +eE_{\Vert
}\left( X-X^{\prime }\right) =0  \label{e1}
\end{equation}%
(here $E_{\Vert }$ is the dc driving electric field directed along the $x$%
-axis) determines the displacement of the electron orbit center, which can be
opposite to the driving force ($X^{\prime }>X$) if $\omega /\omega
_{c}<n^{\prime }-n$. This is the reason for the negative conductivity
effect. It should be noted that in the intersubband displacement
model~\cite{Mon-2011,Mon-2012} the corresponding energy conservation
contains the intersubband excitation energy $%
\Delta _{2}-\Delta _{1}$ instead of $\hbar \omega $.
The displacement model~\cite{Ryz-1969} had been intensively developed after
discovery of ZRS to include self-energy effects and multi-photon processes%
~\cite{DurSac-2003,RyzSur-2003,Ryz-2003}.

Another popular mechanism of MO and negative conductivity effects called
the inelastic model~\cite{DmiVav-2005} originates from oscillations of the electron
distribution function $f\left( \varepsilon \right) $ induced by MW
radiation. An oscillatory behavior of $f\left( \varepsilon \right) $ means
that at certain ranges of $\varepsilon $ there is a population inversion, $%
\partial f\left( \varepsilon \right) /\partial \varepsilon >0$, and the
magnetoconductivity $\sigma _{xx}$ could be negative. These ideas were
also discussed in Ref.~\onlinecite{Dor-2003}.

In this work, we consider mostly the displacement mechanism of MO applied to
the system of SEs in liquid helium: photon-assisted scattering by ripplons.
The inelastic mechanism will be shortly discussed as well, since we see its
certain relation to photon-assisted scattering. As compared to
the initial version of photon-assisted scattering~\cite{BasLev-1965,Ryz-1969},
the theory is extended to include strong Coulomb forces acting between electrons and the
collision broadening of Landau levels (LLs). We found the reason why MO of $%
\sigma _{xx}$ and ZRS caused by photon-assisted scattering
were not seen in experiments on SEs~\cite{KonKon-2009,KonKon-2010}
for chosen ranges of the MW frequency and amplitude, and
formulated conditions under which they could be potentially observed.

\section{Probabilities of photon-assisted scattering}

Consider a 2D electron gas on the free surface of liquid helium in the
presence of a static magnetic field $\mathbf{B}$ directed normally to the
interface. The corresponding vector potential of the magnetic field $\mathbf{%
A}_{0}=\left( 0,Bx,0\right) $. In the presence of the driving electric field
$\mathbf{E}_{\Vert }=(E_{\Vert },0,0)$, eigenfunctions and the energy
spectrum of SEs are characterized by the surface subband number $l$, the
electron orbit center
\begin{equation}
X=-\frac{cp_{y}}{eB}+\zeta ,\text{ \ \ \ }\zeta =-\frac{eE_{\Vert }}{%
m_{e}\omega _{c}^{2}},  \label{e2}
\end{equation}%
and by the LL number $n$:
\begin{equation}
\text{\ }\Psi _{l,n,X}=\psi _{l}\left( z\right) \frac{1}{\sqrt{2\pi \hbar }}%
e^{-i\left( X-\zeta \right) y/l_{B}^{2}}\psi _{n}^{\left( \mathrm{os}\right)
}\left( x-X\right) \text{ ,\ \ }  \label{e3}
\end{equation}%
\begin{equation}
E_{l,n,X}=\Delta _{l}+\hbar \omega _{c}\left( n+1/2\right)
+eE_{\Vert }X+\frac{e^{2}E_{\Vert }^{2}}{2m_{e}\omega _{c}^{2}}\text{ },
\label{e4}
\end{equation}%
Here $\psi _{n}^{\left( \mathrm{os}\right) }\left( x\right) $ are
oscillator eigenfunctions, $\psi _{l}\left( z\right) $ are wavefunctions of
vertical motion, and $l_{B}=\sqrt{\hbar c/eB}$ is the magnetic length.

At low temperatures $T<1\,\mathrm{\ K}$, electrons predominantly occupy the
ground surface subband ($l=1$), since $\Delta _{2}-\Delta _{1}$ is about $6%
\,\mathrm{K}$ (liquid $^{4}\mathrm{He}$) or $3.2\,\mathrm{K}$ (liquid $^{3}%
\mathrm{He}$). The Hamiltonian of electron interaction with MWs has the
usual form \
\begin{equation}
V_{mw}=\frac{e}{cm_{e}}\left( \mathbf{p+}\frac{e}{c}\mathbf{A}_{0 }
\right) \mathbf{\tilde{A}},  \label{e5}
\end{equation}%
where $\mathbf{\tilde{A}}$ is the vector potential of the MW field,
\begin{equation}
\tilde{A}_{\alpha }\left( \mathbf{r}\right) =\sum_{\lambda ,\mathbf{k}%
}\left( \frac{2\pi \hbar c^{2}}{\omega _{k}K_{\mathrm{v}}}\right)
^{1/2}e_{\alpha ,\lambda ,\mathbf{k}}e^{i\mathbf{k}\cdot \mathbf{r}}\left(
a_{\mathbf{k},\lambda }+a_{-\mathbf{k},\lambda }^{\dag }\right) ,  \label{e6}
\end{equation}%
$a_{\mathbf{k},\lambda }$ and $a_{\mathbf{k},\lambda }^{\dag }$ are creation
and destruction operators of photons, $K_{\mathrm{v}}$ is the volume, and $%
\mathbf{e}_{\lambda ,\mathbf{k}}$ is the unit polarization vector. It is
convenient to consider the linear polarization with $\tilde{A}_{y}=0$.

The Hamiltonian of the electron-ripplon interaction, which dominates under
conditions of the experiments ($T=0.2\,\mathrm{K}$)~\cite{KonKon-2010},
can be written as
\begin{equation}
V_{r}=\sum_{\mathbf{q}}V_{r,q}Q_{q}\left( b_{\mathbf{q}}+b_{-\mathbf{q}%
}^{\dag }\right) e^{i\mathbf{q}\cdot \mathbf{r}},  \label{e7}
\end{equation}%
where $b_{\mathbf{q}}$ and $b_{-\mathbf{q}}^{\dag }$ are creation and
destruction operators of ripplons, $V_{r,q}$ is the electron-ripplon coupling%
~\cite{MonKon-book}, $Q_{q}=\sqrt{\hbar q/2\rho \omega _{r,q}D}$, $\omega
_{r,q}\simeq \sqrt{\alpha /\rho }q^{3/2}$, $\alpha $ and $\rho $ are the
surface tension and mass density of liquid helium respectively, and $D$ is
the surface area.

Probabilities of photon-assisted scattering are calculated according to the
Golden Rule with the following matrix elements~\cite{BasLev-1965}
\[
\left\langle i\right\vert \tilde{V}\left\vert f_{\pm }\right\rangle =\sum_{v}%
\frac{\left\langle i\right\vert V_{r}\left\vert v\right\rangle \left\langle
v\right\vert V_{mw}\left\vert f_{\pm }\right\rangle }{E_{i}-E_{v}}+
\]
\begin{equation}
+\sum_{v}%
\frac{\left\langle i\right\vert V_{mw}\left\vert v\right\rangle \left\langle
v\right\vert V_{r}\left\vert f_{\pm }\right\rangle }{E_{i}-E_{v}},
\label{e8}
\end{equation}%
where $E_{i}$ is the energy of the initial state, the final state $%
\left\vert f_{\pm }\right\rangle =\left\vert n^{\prime },X^{\prime },\mathbf{%
n}_{\mathbf{q}}^{\left( r\right) }\pm \mathbf{1}_{\pm \mathbf{q}},\mathbf{n}%
_{\mathbf{k}}^{\left( mw\right) }-\mathbf{1}_{-\mathbf{k}},\right\rangle $
corresponds to processes of destruction of a photon and creation (sign plus)
or destruction (sign minus) of a ripplon. The $\mathbf{n}_{\mathbf{q}%
}^{\left( r\right) }=\left\{ n_{\mathbf{q}}^{\left( r\right) }\right\} $ and
$\mathbf{n}_{\mathbf{k}}^{\left( mw\right) }=\left\{ n_{\mathbf{k}}^{\left(
mw\right) }\right\} $ are the vectors describing occupation numbers of
ripplons and photons.

Further evaluations are based on the relationship $\left( e^{i\mathbf{qr}%
_{e}}\right) _{n,X;n^{\prime },X^{\prime }}=\delta _{X,X^{\prime
}-l_{B}^{2}q_{y}}e^{iq_{x}X}M_{n,n^{\prime }}$ which is valid for the chosen
gauge. Here we use the following notations%
\[
M_{n,n^{\prime }}\left( x_{q},\varphi \right) = i^{\left\vert n^{\prime
}-n\right\vert }J_{n,n^{\prime }}\left( x_{q}\right) \times
\]
\begin{equation}
\times
\exp \left[
ix_{q}\cos \left( \varphi \right) \sin \left( \varphi \right) +i\left(
n^{\prime }-n\right) \varphi \right],
\label{e9}
\end{equation}%
\[
J_{n,n^{\prime }}\left( x_{q}\right) =\sqrt{\frac{\min \left( n,n^{\prime
}\right) !}{\max \left( n,n^{\prime }\right) !}}x_{q}^{\frac{\left\vert
n^{\prime }-n\right\vert }{2}} \times
\]
\begin{equation}
\times \exp \left( -\frac{x_{q}}{2}\right) L_{\min
\left( n,n^{\prime }\right) }^{\left\vert n^{\prime }-n\right\vert }\left(
x_{q}\right) ,  \label{e10}
\end{equation}%
where $x_{q}=q^{2}l_{B}^{2}/2$ is a dimensionless parameter, $q_{x}=q\cos
\varphi $, and $L_{n}^{m}(x)$ are the associated Laguerre polynomials. Our
expression for $M_{n,n^{\prime }}\left( x_{q},\varphi \right) $ differs
from the similar equation of Ref.~\onlinecite{BasLev-1965} because we have chosen
the different gauge. The result of Ref.~\onlinecite{BasLev-1965} can be restored
from Eq.~(\ref{e9}) using the substitution $\varphi \rightarrow \varphi -\pi /2$.

Introducing%
\begin{equation}
\mathcal{E}\left( \omega \right) =\sqrt{4\pi \frac{n^{\left( mw\right)
}\left( \omega \right) }{K_{\mathrm{v}}}\hbar \omega }  \label{e11}
\end{equation}%
and \ $C _{\mathbf{q}}^{\left( \pm \right) }=V_{r,q}Q_{q}\left[ n_{\pm
\mathbf{q}}^{\left( r\right) }+\frac{1}{2}\pm \frac{1}{2}\right] ^{1/2}$,
and following the procedure described in Ref.~\onlinecite{BasLev-1965}, one can find%
\[
\left\vert \left\langle i\right\vert \tilde{V}\left\vert f_{\pm }
\right\rangle \right\vert =C_{\mathbf{q}}^{\left( \pm \right) }\frac{%
e\mathcal{E}\left( \omega \right) }{2m_{e}\omega ^{2}l_{B}}\delta
_{X,X^{\prime }-l_{B}^{2}q_{y}}\times
\]%
\begin{equation}
\sum_{n^{\prime \prime }}\left[ \frac{\omega M_{n,n^{\prime \prime
}}p_{n^{\prime \prime },n^{\prime }}}{\omega _{c }\left( n-n^{\prime
\prime }\right) -q_{y}V_{H}\mp \omega _{r,q}}+\frac{\omega p_{n,n^{\prime
\prime }}M_{n^{\prime \prime },n^{\prime }}}{\omega _{c} \left(
n-n^{\prime \prime }\right) +\omega }\right],  \label{e12}
\end{equation}%
where $p_{n,n^{\prime }}=i\left( \sqrt{n}\delta _{n^{\prime },n-1}-\sqrt{%
n^{\prime }}\delta _{n^{\prime },n+1}\right) $ are dimensionless matrix
elements of the electron momentum operator, and $V_{H}$ is the absolute
value of the Hall velocity: $eE_{\Vert }\left( X-X^{\prime }\right) =-\hbar
q_{y}V_{H}$.

Keeping in mind the energy conservation delta-function, in the first term
of Eq.~(\ref{e12}) we
can use the replacement $\omega _{c}\left( n-n^{\prime \prime
}\right) -q_{y}V_{H}\mp \omega _{r,q}\rightarrow \omega _{c}\left( n^{\prime
}-n^{\prime \prime }\right) -\omega $. For slightly broadened LLs, this is
an approximate procedure\ which has the same accuracy as the replacement $%
1-f\left( \varepsilon ^{\prime }\right) \rightarrow 1-f\left( \varepsilon
\right) $ in usual conductivity equations. Then, using properties of the
associated Laguerre polynomials, one can find
\[
\left\vert \left\langle i\right\vert \tilde{V}\left\vert f_{\pm
}\right\rangle \right\vert ^{2}=\left( C _{\mathbf{q}}^{\left( \pm
\right) }\right) ^{2}\left[ \frac{e\mathcal{E}\left( \omega \right) }{%
2m_{e}\omega ^{2}l_{B}}\right] ^{2} \times
\]
\begin{equation}
\times \delta _{X,X^{\prime
}-l_{B}^{2}q_{y}}\chi _{\mathbf{q}}\left( \omega \right) x_{q}J_{n,n^{\prime
}}^{2}\left( x_{q}\right) ,  \label{e13}
\end{equation}%
where%
\begin{equation}
\chi _{\mathbf{q}}\left( \omega \right) =\left\vert \frac{e^{i\varphi
}\omega }{\omega _{c}+\omega }-\frac{e^{-i\varphi }\omega }{\omega
_{c}-\omega }\right\vert ^{2}.  \label{e14}
\end{equation}%
The dependence of $\chi _{\mathbf{q}}$ on $\varphi $ is important for
calculation of the momentum relaxation rate, which contains an additional
factor $q_{y}^{2}=q^{2}\sin ^{2}\varphi $.

After summation over $\mathbf{k}$ the probabilities of photon-assisted
scattering from $n,X$ to $n^{\prime },X^{\prime }$ accompanied by the momentum
exchange $\hbar \mathbf{q}$ can be found as
\[
w_{n,X\rightarrow n^{\prime },X^{\prime }}^{\left( \pm \right) }\left(
\mathbf{q}\right) =\frac{2\pi }{\hbar }\left( C _{\mathbf{q}}^{\left(
\pm \right) }\right) ^{2}\lambda _{mw}\chi _{\mathbf{q}}\left( \omega
\right) x_{q}\times
\]%
\begin{equation}
\times \delta _{X,X^{\prime }-l_{B}^{2}q_{y}}J_{n,n^{\prime }}^{2}\left(
x_{q}\right) \delta (\varepsilon _{n}-\varepsilon _{n^{\prime }}-\hbar
q_{y}V_{H}\mp \hbar \omega _{q}+\hbar \omega ),  \label{e15}
\end{equation}%
where $\varepsilon _{n}=\hbar \omega _{c}\left( n+1/2\right)$,%
\begin{equation}
\lambda _{mw}=\frac{e^{2}E_{mw}^{2}}{4m_{e}^{2}\omega ^{4}l_{B}^{2}},\text{
\ }E_{mw}^{2}=4\pi \frac{N_{mw}\left( \omega \right) }{K_{\mathrm{v}}}\hbar
\omega ,  \label{e16}
\end{equation}%
$N_{mw}\left( \omega \right) $ is the number of photons with the frequency $%
\omega $, and $E_{mw}$ ($\sqrt{\mathrm{ergs}/\mathrm{cm}^{3}}$) is the
amplitude of the electric field in the MW. As compared to usual electron-ripplon
scattering, Eq.~(\ref{e15}) contains the photon energy $\hbar \omega $ in the
argument of the delta-function and additional dimensionless proportionality
factors $\lambda _{mw}$, $\chi _{\mathbf{q}}\left( \omega \right) $, and $%
x_{q}$. The $\chi _{\mathbf{q}}\left( \omega \right) $ is of
the order of unity (here $\omega $ is substantially larger than $\omega _{c}$%
), while $x_{q}$ is of the order of $n'-n$ due to $J_{n,n^{\prime }}^{2}\left( x_{q}\right)$.
The $\lambda _{mw}$ depends only on the MW field parameters ($E_{mw}$
and $\omega $) and on basic properties of the electron gas under magnetic
field.

\section{The conductivity of strongly interacting electrons}

Generally, the structure of Eq.~(\ref{e15}) is similar to the structure of
the corresponding probability of the usual electron-ripplon scattering.
Therefore, considering the contribution of photon-assisted scattering into
the momentum relaxation rate, we can use advantages of the
approach~\cite{MonKon-book,MonTesWyd-2002,Mon-2013},
which allows to express average scattering probabilities and the
momentum relaxation rate $\nu _{\mathrm{eff}}$ in terms of the dynamic
structure factor (DSF) of the 2D electron system. Since $\sum_{X^{\prime
}}w_{n,X\rightarrow n^{\prime },X^{\prime }}^{\left( \pm \right) }\left(
\mathbf{q}\right) $ does not depend on $X$, when averaging over the initial
electron states we can consider electron distribution over LLs only $%
f_{n}\simeq Z_{\Vert }^{-1}e^{-\varepsilon _{n}/T_{e}}$ (here $Z_{\Vert
}=\sum_{n}e^{-\varepsilon _{n}/T_{e}}$). Then, the average probability of
electron scattering with the momentum exchange $\hbar \mathbf{q}$
caused by destruction of a photon and creation or destruction of a
ripplon can be written in the following form
\[
\bar{w}_{\mathrm{mw},\mathbf{q}}^{\left( \pm \right) }=\frac{1}{\hbar ^{2}}%
\left( C _{r,\mathbf{q}}^{\left( \pm \right) }\right) ^{2}\lambda
_{mw}\chi _{\mathbf{q}}\left( \omega \right) \times
\]
\begin{equation}
\times x_{q}S\left( q,-q_{y}V_{H}\mp
\omega _{r,q}+\omega \right) ,  \label{e17}
\end{equation}%
where $C _{r,\mathbf{q}}^{\left( \pm \right) }$ is obtained from
$C _{\mathbf{q}}^{\left( \pm \right) }$ replacing the ripplon
occupation number $n_{\pm \mathbf{q}}^{\left( r\right) }$ with the
Bose distribution function $N_{r,\pm \mathbf{q}}$, the function
\[
S\left( q,\Omega \right) =\frac{2}{\pi \hbar Z_{\Vert }}\sum_{n,n^{\prime
}}J_{n,n^{\prime }}^{2}\left( x_{q}\right) \times
\]
\begin{equation}
\times \int d\varepsilon e^{-\varepsilon
/T_{e}}g_{n}\left( \varepsilon \right) g_{n^{\prime }}\left( \varepsilon
+\hbar \Omega \right)   \label{e18}
\end{equation}%
is the DSF of a nondegenerate 2D electron gas, $g_{n}\left( \varepsilon
\right) =-\mathrm{Im}G_{n}\left( \varepsilon \right) $, and $G_{n}\left(
\varepsilon \right) $ is the single-electron Green's function.

Taking into account the collision broadening of LLs, we shall use the result
of the cumulant approach~\cite{Ger-1976}
\begin{equation}
g_{n}\left( \varepsilon \right) =\frac{\sqrt{2\pi }\hbar }{\Gamma _{n}}\exp %
\left[ -\frac{2\left( \varepsilon -\varepsilon _{n}\right) ^{2}}{\Gamma
_{n}^{2}}\right] ,  \label{e19}
\end{equation}%
where $\Gamma _{n}$ is the collision broadening of LLs~\cite{AndUem-1974}. This
approximation, being quite accurate for low LLs, greatly simplifies further
evaluations. The equilibrium DSF has an important property
\begin{equation}
S\left( q,-\Omega \right) =e^{-\hbar \Omega /T_{e}}S\left( q,\Omega \right)
\label{e20}
\end{equation}%
which allows to shorten evaluations. According to Eqs.~(\ref{e18}) and (\ref{e19}),
the $S\left( q,\Omega \right) $ as a function of frequency has
sharp maxima when $\Omega $ approaches the LL excitation frequency $\left(
n^{\prime }-n\right) \omega _{c}$.

The momentum relaxation rate can be found by evaluating the total momentum
gained by scatterers. For an infinite isotropic system, the kinetic
friction acting on the electron gas,
\begin{equation}
\mathbf{F}_{\mathrm{fric}}=-N_{e}\sum_{\mathbf{q}}\hbar \mathbf{q}\left(
\bar{w}_{r,\mathbf{q}}+\bar{w}_{\mathrm{mw},\mathbf{q}}^{\left( +\right) }%
\mathbf{+}\bar{w}_{\mathrm{mw},\mathbf{q}}^{\left( -\right) }\right) ,
\label{e21}
\end{equation}%
should be antiparallel to the current. Here $\bar{w}_{r,\mathbf{q}}\equiv
\bar{w}_{r,\mathbf{q}}^{\left( +\right) }+\bar{w}_{r,\mathbf{q}}^{\left(
-\right) }$ is the corresponding probability obtained for usual
electron-ripplon scattering in the absence of MW radiation, and $\bar{w}_{r,%
\mathbf{q}}^{\left( \pm \right) }$ can be found from Eq.~(\ref{e17}) using
the substitution $\lambda _{mw}\chi _{\mathbf{q}}\left( \omega \right)
x_{q}\rightarrow 1$ and setting $\omega \rightarrow 0$ in the frequency
argument of the DSF.

Thus, the momentum relaxation rate $\nu
_{\mathrm{eff}}$ can be defined by the relationship $\left( \mathbf{F}_{\mathrm{fric}%
}\right) _{y}=-N_{e}m_{e}\nu _{\mathrm{eff}}V_{y}$ , where \ $V_{y}\simeq -V_{H}$.
Using elastic approximation ($\hbar
\omega _{r,q}\rightarrow 0$), the correction into the effective collision
frequency induced by MW radiation is found as
\begin{equation}
\nu _{\mathrm{mw}}=\frac{2\lambda _{mw}}{m_{e}\hbar }\sum_{\mathbf{q}%
}q_{y}^{2}\chi _{\mathbf{q}}\left( \omega \right) x_{q} C _{r,\mathbf{q}%
}^{2}S^{\prime }\left( q,\omega \right) ,  \label{e22}
\end{equation}%
where $C _{r,\mathbf{q}}\equiv C _{r,\mathbf{q}}^{\left( -\right)
}\simeq C _{r,\mathbf{q}}^{\left( +\right) }$, and $S^{\prime }\left(
q,\Omega \right) =\partial S\left( q,\Omega \right) /\partial \Omega $.

In the absence of MW radiation, the momentum relaxation rate $\nu
_{r}^{\left( 0\right) }$ can be formally obtained from Eq.~(\ref%
{e22}) using the replacement $\lambda _{mw}\chi _{\mathbf{q}}\left( \omega
\right) x_{q}\rightarrow 1$ and setting $\omega \rightarrow 0$ in the frequency
argument of $S^{\prime }$:%
\begin{equation}
\nu _{r}^{\left( 0\right) }=\frac{2}{m_{e}\hbar }\sum_{\mathbf{q}}q_{y}^{2}%
C _{r,\mathbf{q}}^{2}S^{\prime }\left( q,0\right) .  \label{e23}
\end{equation}%
The derivative of the Eq.~(\ref{e20}) gives the relationship%
\begin{equation}
S^{\prime }\left( q,0\right) =\frac{\hbar }{2T_{e}}S\left( q,0\right) >0,
\label{e24}
\end{equation}%
which transforms $\nu _{r}^{\left( 0\right) }$ into the result of
the SCBA theory~\cite{AndUem-1974} applied to the system of SEs on liquid helium\cite%
{MonKon-book}. The total momentum relaxation rate $\nu _{\mathrm{eff}}=\nu
_{r}^{\left( 0\right) }+\nu _{\mathrm{mw}}$.

The results of Eqs.~(\ref{e22}) and (\ref{e23}) can be obtained also using
the direct definition of the electron current
\begin{equation}
j_{x}=-en_{s}\sum_{\mathbf{q}}\left( X^{\prime }-X\right) _{\mathbf{q}%
}\left( \bar{w}_{r,\mathbf{q}}+\bar{w}_{\mathrm{mw},\mathbf{q}}^{\left(
+\right) }+\bar{w}_{\mathrm{mw},\mathbf{q}}^{\left( -\right) }\right).
\label{e25}
\end{equation}%
Taking into account $\left( X^{\prime }-X\right) _{\mathbf{q}}=q_{y}l_{B}^{2}
$, one can find%
\[
\sigma _{xx}=\frac{en_{s}l_{B}^{2}}{E_{\Vert }}\sum_{\mathbf{q}}q_{y}\left(
\bar{w}_{r,\mathbf{q}}+\bar{w}_{\mathrm{mw},\mathbf{q}}^{\left( +\right) }+%
\bar{w}_{\mathrm{mw},\mathbf{q}}^{\left( -\right) }\right) \simeq
\]%
\begin{equation}
\simeq \frac{e^{2}n_{s}}{m_{e}}\frac{\nu _{r}^{\left( 0\right) }+\nu _{%
\mathrm{mw}}}{\omega _{c}^{2}},  \label{e26}
\end{equation}%
which proves that $\mathbf{F}_{\mathrm{fric}}$ is antiparallel to the
current. It should be noted that one cannot disregard the driving field
correction $-q_{y}V_{H}$ in the expressions for $\bar{w}_{\mathrm{mw},%
\mathbf{q}}^{\left( \pm \right) }$ and $\bar{w}_{r,\mathbf{q}}$, otherwise
scattering probabilities in the direction of the driving force and in the
opposite direction would be the same leading to $j_{x}=0$.

SEs on liquid helium form a highly correlated 2D electron
system. Even for small densities $n_{e}\sim 10^{6}\,\mathrm{cm}^{-2}$, the
average energy of Coulomb interaction per particle is much larger than the
average kinetic energy ($T_{e}$). Under such conditions, the 2D electron
system subjected to a magnetic field can be well described using the concept
of the fluctuational electric field~\cite{DykKha-1979}. At a finite temperature,
for each electron there is an electric field $\mathbf{E}%
_{f}$ of other electrons caused by fluctuations and
directed to an equilibrium position of the
electron. Within the cyclotron orbit length, the $\mathbf{E}_{f}$ can be
approximately considered as a uniform field, if $B$ is strong enough. Thus,
the magnetic field and $\mathbf{E}_{f}$ cause fast rotation of the electron obit
center around the equilibrium position. Therefore, such a strongly
interacting system can be considered as an ensemble of noninteracting
electrons, whose orbit centers move fast in the fluctuational field\cite%
{MonTesWyd-2002,MonKon-book}. The distribution of $E_{f}$ is known from numerical
calculations~\cite{FanDykLea-1997}.

The fluctuational electric field introduces an additional broadening of
maxima of the DSF~\cite{MonKon-book,Mon-2012}:
\begin{equation}
S\left( q,\Omega \right) =\frac{2\sqrt{\pi }}{Z_{\parallel }}%
\sum_{n,n^{\prime }}\frac{J_{n,n^{\prime }}^{2}}{\gamma _{n,n^{\prime }}}%
\exp \left[ -\frac{\varepsilon _{n}}{T_{e}}-P_{n,n^{\prime }}\left( \Omega
\right) \right] \text{ },  \label{e27}
\end{equation}%
where%
\begin{equation}
P_{n,n^{\prime }}=\frac{\left[ \Omega -\left( n^{\prime }-n\right) \omega
_{c}-\phi _{n}\right] ^{2}}{\gamma _{n,n^{\prime }}^{2}},\text{ }\phi _{n}=%
\frac{\Gamma _{n}^{2}+x_{q}\Gamma _{C}^{2}}{4T_{e}\hbar },  \label{e28}
\end{equation}%
and
\begin{equation}
\hbar \gamma _{n,n^{\prime }}=\sqrt{\frac{\Gamma _{n}^{2}+\Gamma _{n^{\prime
}}^{2}}{2}+x_{q}\Gamma _{C}^{2}}.  \label{e29}
\end{equation}%
In Eq.~(\ref{e28}), defining $P_{n,n^{\prime }}\left( \Omega \right) $, we have
neglected terms of the order of $\Gamma _{n}^{2}/8T_{e}^{2}$ which are very
small for the considered system. The Coulomb broadening parameter $\Gamma
_{C}=\sqrt{2}eE_{f}^{(0)}l_{B}$ increases with $n_{e}$ and $T_{e}$ because $%
E_{f}^{(0)}\simeq 3\sqrt{T_{e}}n_{e}^{3/4}$.

The fluctuational field introduces also an additional shift in positions of
maxima of the DSF $\phi _{C}=x_{q}\Gamma _{C}^{2}/4T_{e}\hbar $ which enters
the definition of $\phi _{n}$. This Coulomb shift is restored from the
expression for the DSF of the 2D Wigner solid under a strong magnetic field.
It preserves the equilibrium property of Eq.~(\ref{e20}). The influence of
this shift on the intersubband displacement mechanism of MO~\cite{Mon-2012}
was recently confirmed in experiments~\cite{KonMonKon-2013}
on SEs above liquid $^{3}\mathrm{He}$.

In Eq.~(\ref{e22}), describing $\nu _{\mathrm{mw}}$, the function $\chi _{\mathbf{q}%
}\left( \omega \right) $ is averaged over directions of the ripplon
vector $\left\langle \chi _{\mathbf{q}}\left( \omega \right) \sin
^{2}\varphi \right\rangle _{\varphi }\equiv \chi _{\mathrm{tr}}\left( \omega
\right) $. Simple integration yields
\begin{equation}
\chi _{\mathrm{tr}}\left( \omega \right) =\frac{1}{2}\frac{3\omega
_{c}^{2}/\omega ^{2}+1}{\left( \omega _{c}^{2}/\omega ^{2}-1\right) ^{2}}.
\label{e30}
\end{equation}%
For $\omega ^{2}\gg 3\omega _{c}^{2}$, the $\chi _{\mathrm{tr}}\left( \omega
\right) \rightarrow 1/2$ which coincides with $\left\langle \sin ^{2}\varphi
\right\rangle _{\varphi }$ entering $\nu _{r}^{\left( 0\right) }$. Using
this notation, the photon-assisted scattering correction $\nu _{\mathrm{mw}}$
can be represented in an analytical form
\begin{equation}
\nu _{\mathrm{mw}}=\lambda _{mw}\chi _{\mathrm{tr}}\left( \omega \right)
\frac{2m_{e}T\Lambda ^{2}}{\sqrt{\pi }\alpha \hbar ^{3}l_{B}^{2}}F_{\omega
}\left( B\right) ,  \label{e31}
\end{equation}%
where%
\[
F_{\omega }\left( B\right) =-\frac{1}{Z_{\Vert }}\sum_{n,m}e^{-\varepsilon
_{n}/T_{e}}\int_{0}^{\infty }dx_{q}x_{q}V_{1,1}^{2}\left( x_{q}\right)
J_{n,n+m}^{2}\times
\]%
\begin{equation}
\frac{\omega /\omega _{c}-m-\phi _{n}/\omega _{c}}{\tilde{\gamma}_{n,n+m}^{3}%
}\exp \left[ -\left( \frac{\omega /\omega _{c}-m-\phi _{n}/\omega _{c}}{%
\tilde{\gamma}_{n,n+m}}\right) ^{2}\right] ,  \label{e32}
\end{equation}%
$\tilde{\gamma}_{n,n+m}=\gamma _{n,n+m}/\omega _{c}$, and $m=n^{\prime }-n$.
The dimensionless electron-ripplon coupling~\cite{MonKon-book}
\begin{equation}
V_{1,1}\left( x_{q}\right) =x_{q}w_{c}\left( x_{q}/2g^{2}l_{B}^{2}\right)
+eE_{\bot }l_{B}^{2}/\Lambda   \label{e33}
\end{equation}%
is defined by the function%
\begin{equation}
w_{c}\left( x\right) =-\frac{1}{1-x}+\frac{1}{\left( 1-x\right) ^{3/2}}\ln
\left( \frac{1+\sqrt{1-x}}{\sqrt{x}}\right) .  \label{e34}
\end{equation}%
The $V_{1,1}\left( x\right) $ depends also on the pressing electric field $%
E_{\bot }$, the image potential parameter $\Lambda =e^{2}\left( \epsilon
-1\right) /4\left( \epsilon +1\right) $ (here $\epsilon $ is the dielectric
constant of liquid helium), and on the localization parameter ($g$) of the
SE wave function: $\psi _{1}\left( z\right) =2g^{3/2}z\exp (-gz)$.

\begin{figure}[tbp]
\begin{center}
\includegraphics[width=10.6cm]{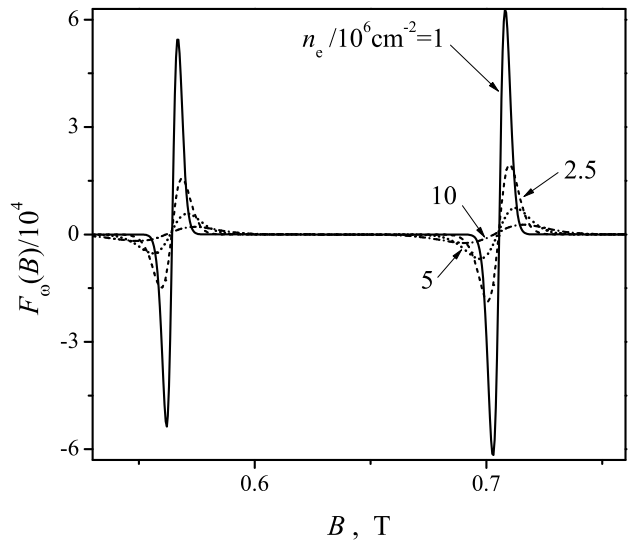}
\end{center}
\caption{$F_{\omega }(B)$ calculated near $\omega/\omega_{c}=4$
and $5$ for $T=0.2\,\mathrm{K}$ and
four electron densities: $n_{e}= 1\cdot 10^{6}\,\mathrm{cm}^{-2}$
(solid), $2.5\cdot 10^{6}\,\mathrm{cm}^{-2} $ (dashed),
$5\cdot 10^{6}\,\mathrm{cm}^{-2} $ (dotted), $10\cdot 10^{6}\,
\mathrm{cm}^{-2} $ (dash-dotted).} \label{f1}
\end{figure}

The absolute value of the function $F_{\omega
}\left( B\right) $ defined by Eq.~(\ref{e32}) increases
sharply when $\omega /\omega _{c}$ becomes
close to an integer $m=n^{\prime }-n>0$. In the vicinity of this point,
$F_{\omega }\left( B\right) $ changes its sign because of the factor $%
\omega /\omega _{c}-m-\phi _{n}/\omega _{c}$. The typical dependence $%
F_{\omega }\left( B\right) $ is shown in Fig.~\ref{f1} for four different
electron densities. We have chosen the MW frequency
$\omega /2\pi =79\,\mathrm{GHz}$ which is typical for SEs above liquid
$^{3}\mathrm{He}$.  It is obvious that an increase in $n_{e}$ suppresses the
amplitude of MO and broadens their shape. As expected, negative values of $%
F_{\omega }\left( B\right) $ and $\nu _{\mathrm{mw}}$ mostly appear at $%
\omega /\omega _{c}\left( B\right) >m$.

\section{Discussion and conclusions}

Negative conductivity effects and ZRS appear when $\nu _{\mathrm{mw}}\gtrsim
\nu _{\mathrm{eff}}^{\left( 0\right) }$. It should be notes that $\chi _{%
\mathrm{tr}}\left( \omega \right) $ is about $1/2$, if $\omega $ is
substantially larger than $\omega _{c}$. As compared to the result of usual
electron-ripplon scattering, the momentum relaxation rate caused by
photon-assisted scattering [Eq.~(\ref{e31})] has an important proportionality
factor $\lambda _{mw}$, which originates from $V_{mw}^{2}/\left(
E_{i}-E_{v}\right) ^{2}\sim V_{mw}^{2}/\hbar ^{2}\omega ^{2}$. We note also
the presence of the MW frequency $\omega $ in expression for $F_{\omega
}\left( B\right) $ given in Eq.~(\ref{e32}). For usual scattering, the main
contribution into $F_{0}\left( B\right) $ comes from terms with $m=0$, when
the factor $\omega /\omega _{c}-m-\phi _{n}/\omega _{c}\rightarrow -\phi
_{n}/\omega _{c}$ and $F_{0}\left( B\right) >0$. In the case of
photon-assisted scattering with $\omega >\omega _{c}$, the extrema of $%
F_{\omega }\left( B\right) $ occur at $\left\vert \omega /\omega
_{c}-m\right\vert \sim \tilde{\gamma}_{n,n+m}$. Therefore, the ratio $%
F_{\omega }\left( B\right) /F_{0}\left( B\right) \approx $ $4T_{e}/\hbar
\gamma _{n,n}$, and we can roughly estimate
\begin{equation}
\frac{\nu _{\mathrm{mw}}}{\nu _{r}^{\left( 0\right) }}\sim \lambda _{mw}%
\frac{4T_{e}}{\sqrt{\Gamma _{n}^{2}+\Gamma _{C}^{2}}}=\frac{%
e^{2}E_{mw}^{2}\omega _{c}T_{e}}{m_{e}\hbar \omega ^{4}\sqrt{\Gamma
_{n}^{2}+\Gamma _{C}^{2}}}.  \label{e35}
\end{equation}%
The $\nu _{\mathrm{mw}}/\nu _{r}^{\left( 0\right) }$ increases with the MW
field amplitude $E_{mw}$ and decreases with the MW frequency.

A more accurate comparison of $\nu _{\mathrm{mw}}$ and $\nu _{r}^{\left(
0\right) }$ can be given using numerical evaluation of Eq.~(\ref{e32}) shown
in Fig.~\ref{f1}. Consider typical conditions of the experiment~\cite{KonKon-2010} with
SEs on liquid $^{3}\mathrm{He}$:   $%
\omega /2\pi \simeq 79\,\mathrm{GHz}$, $T=0.2\,K$ and $n_{e}=10^{6}\,\mathrm{cm}^{-2}$.
Under these conditions, the intersubband displacement
mechanism~\cite{Mon-2012} leads to giant MO and $\sigma _{xx}<0$ already at a MW field
amplitude which corresponds to the Rabi frequency $eE_{mw}\left\vert
\left\langle 1\right\vert z\left\vert 2\right\rangle \right\vert /\hbar
=10^{8}\,\mathrm{s}^{-1}$. For such a MW power, $E_{mw}\simeq 0.1\,\mathrm{V/cm}$%
, and the calculation based on Eqs.~(\ref{e31}) and (\ref{e32}) gives $\delta
\nu _{\mathrm{eff}}/\nu _{\mathrm{eff}}^{\left( 0\right) }\simeq 0.65\cdot
10^{-5}$, if $B$ \ is close to $0.7\,\mathrm{T}$ ($m=4$). This explains why MO
caused by photon-assisted scattering were not observed together with MO
caused by the intersubband displacement mechanism.

The correction to $\nu _{r}^{(0)}$ caused by the intersubband
displacement mechanism~\cite{Mon-2012} does not have the proportionality
factor $\lambda _{mw}$, because this mechanism does not involve photons.
Instead, there is the proportionality factor $[\bar{n}_{2}-e^{-\left( \Delta
_{2}-\Delta _{1}\right) /T_{e}}\bar{n}_{1}]$, where $\bar{n}_{l}=N_{l}/N_{%
\mathrm{all}}$ is the fractional occupancy of the $l$-th surface subband.
This factor is not too small, typically about $0.1$ or even larger.
Therefore, the enhancement factor $4T_{e}/\hbar \gamma _{n,n^{\prime }}$ noted above
can make the amplitude of oscillations very large.

Thus, the main reason why photon-assisted scattering is small in the system
of SEs on liquid helium is the parameter $\lambda _{mw}$, which actually
does not depend on the nature of scatterers. For a fixed ratio $\omega
/\omega _{c}$, it depends only on parameters of the MW field ($E_{mw}$ and $%
\omega $) and on basic parameters of charge carriers (the charge and the
effective mass $m_{e}^{\ast }$). According to recent treatments of
photon-assisted scattering in semiconductor 2D electron systems~\cite{Ryz-2003},
the effect of MW on the in-plane current $j_{x}$ is
characterized by the factor $J_{M}^{2}\left( \sqrt{2x_{q}}E_{mw}/\mathcal{%
\tilde{E}}_{\omega }\right) $, where $J_{M}\left( z\right) $ is the Bessel
function, $M$ is the number of photons assisted in a scattering event, and $%
\mathcal{\tilde{E}}_{\omega }$ is a characteristic MW field%
\begin{equation}
\mathcal{\tilde{E}}_{\omega }=\frac{\sqrt{2m_{e}^{\ast }}\omega \left\vert
\omega _{c}^{2}-\omega ^{2}\right\vert \hbar ^{1/2}}{e\sqrt{\omega
_{c}^{2}+\omega ^{2}}\omega _{c}^{1/2}}  \label{e36}
\end{equation}%
For $\omega ^{2}\gg \omega _{c}^{2}$, the $\mathcal{\tilde{E}}_{\omega
}\simeq \sqrt{2}m_{e}^{\ast }\omega ^{2}l_{B}/e$, and the argument of the
Bessel function $\sqrt{2x_{q}}E_{mw}/\mathcal{\tilde{E}}_{\omega
}\rightarrow 2\sqrt{x_{q}\lambda _{mw}}$. In the limit of weak MW
fields $E_{mw}\ll \mathcal{\tilde{E}}_{\omega }$, the effect of
one-photon assisted scattering is characterized by the small proportionality
factor $J_{1}^{2}\left( 2\sqrt{x_{q}\lambda _{mw}}\right) \simeq
x_{q}\lambda _{mw}$, which agrees with our calculations.

The main peculiarity of the electron gas in GaAs/AlGaAs, as compared to SEs on
liquid helium, is the small effective mass $m_{e}^{\ast }\simeq 0.067\,m_{e}$.
Therefore, for the fixed ratio $\omega /\omega _{c}=2$, the Eq.~(\ref{e36})
yields: $\mathcal{\tilde{E}}_{\omega }\simeq 74\,\mathrm{V/cm}$ if $\omega /2\pi =79%
\,\mathrm{GHz}$, and $\mathcal{\tilde{E}}_{\omega }\simeq 22\,\mathrm{V/cm}$ if $%
\omega /2\pi =35.5\,\mathrm{GHz}$. It should be noted that these values of $\mathcal{%
\tilde{E}}_{\omega }$ are much larger than the estimate of the MW field $%
E_{mw}\approx 0.5\,\mathrm{V/cm}$ given in Ref.~\onlinecite{RivSch-2004} for typical
experimental conditions realized in semiconductor systems. For SEs on liquid
helium the estimate of the characteristic MW field $\mathcal{\tilde{E}}%
_{\omega }$ increases only by the factor $1/\sqrt{0.067}\simeq 3.9$. This
factor could be compensated by coordinated reductions in the MW frequency
and the cyclotron frequency provided the ratio $\omega _{c}/\omega $ and the
MW power are fixed. Therefore, if photon-assisted scattering is the main origin
of MO in GaAs/AlGaAs, it could be potentially observed also in the system of
SEs on liquid helium in the low MW frequency range and at highest radiation
power.

At this point it is instructive to discuss briefly the inelastic mechanism
of MO and ZRS, which is expected to produce larger MO amplitudes than those of the
displacement mechanism. In the inelastic model~\cite{DmiVav-2005}, MO and the negative
conductivity effect originate from oscillatory behavior of the electron
distribution function $f\left( \varepsilon \right) $ entering the conductivity
equation for usual scattering which does not involve photons. Oscillatory corrections
$\delta f$ to the equilibrium distribution function $f_{F}\left( \varepsilon \right) $
are caused by the MW field. The important point of the model is the
relationship between $\sigma _{xx}$ and $\partial f\left( \varepsilon
\right) /\partial \varepsilon $. Sometimes this relationship is referred to
Kubo conductivity equations. It should be noted that there the factor $%
\partial f\left( \varepsilon \right) /\partial \varepsilon $ appears from
the property of the equilibrium distribution function $f\left( \varepsilon
\right) [1-f\left( \varepsilon \right) ]=-T_{e}\partial f/\partial \varepsilon $%
, which is not valid for an arbitrary function [moreover $f\left( 1-f\right)
>0$]. The proper relationship between $\nu _{r}^{\left( 0\right) }$ and $%
\partial f/\partial \varepsilon $ can be found considering the
nonequilibrium DSF of a 2D electron gas%
\[
S\left( q,\Omega \right) =\frac{D}{N_{e}\pi ^{2}l_{B}^{2}\hbar }%
\sum_{n,n^{\prime }}J_{n,n^{\prime }}^{2}\left( x_{q}\right) \int
d\varepsilon f\left( \varepsilon \right) \times
\]%
\begin{equation}
\times \left[ 1-f\left( \varepsilon +\hbar \Omega \right) \right]
g_{n}\left( \varepsilon \right) g_{n^{\prime }}\left( \varepsilon +\hbar
\Omega \right) ,  \label{e37}
\end{equation}%
where $f\left( \varepsilon \right) $ is an arbitrary distribution function and
$D$ is the surface area. Now the
$S^{\prime }\left( q,0\right) \neq \hbar S\left( q,0\right) /2T_{e}$
and one cannot guarantee that $S^{\prime }\left( q,0\right) $ entering Eq.~(%
\ref{e23}) is positive.

Generally, $S^{\prime }\left( q,\Omega \rightarrow 0\right) $ consist of the
term with the derivative of the factor $1-f\left( \varepsilon \right) $ and
the term with the derivative of $g_{n^{\prime }}\left( \varepsilon \right)
$. The second term can be rearranged using integration by parts. Then, using
the property $J_{n,n^{\prime }}=J_{n^{\prime },n}$ one can find%
\[
S^{\prime }\left( q,0\right) =\frac{D}{N_{e}2\pi ^{2}l_{B}^{2}}%
\sum_{n,n^{\prime }}J_{n,n^{\prime }}^{2}\left( x_{q}\right) \times
\]
\begin{equation}
\times \int
d\varepsilon \left[ -\frac{\partial f\left( \varepsilon \right) }{\partial
\varepsilon }\right] g_{n}\left( \varepsilon \right) g_{n^{\prime
}}\left( \varepsilon \right) .  \label{e38}
\end{equation}%
This equation together with Eq.~(\ref{e23}) establishes the necessary
relationship between the momentum relaxation rate and $\partial f\left(
\varepsilon \right) /\partial \varepsilon $. For interaction with
short-range scatterers like vapor atoms, the coupling parameter
$C _{\mathbf{q}}^{2}= \hbar ^{3}\nu _{0}/2m_{e}D
$, where $\nu _{0}$ is the collision frequency at $B=0$.

To find corrections to the equilibrium distribution function induced by MW
radiation, a sort of kinetic equation was used~\cite{DmiVav-2005}
where the effect of microwaves $%
\mathrm{St}_{mw}\left\{ f\left( \varepsilon \right) \right\} $ was
proportional to $f\left( \varepsilon \pm \hbar \omega \right) -f\left(
\varepsilon \right) $. Since the MW itself cannot cause electron transitions
from $\varepsilon $ to $\varepsilon \pm \hbar \omega $ when $\omega /\omega
_{c}\geq 2$, we conclude that the main contribution into $\mathrm{St}%
_{mw}\left\{ f\left( \varepsilon \right) \right\} $ comes from
photon-assisted scattering. Though, inelastic mechanism applied to SEs on
liquid helium requires a separate investigation, we expect that oscillatory
corrections to the equilibrium distribution function will contain the small
parameter $\lambda _{mw}$ introduced above. Additionally, electron-electron
interaction, which is extremely strong for SEs on liquid helium, should
increase substantially the inelastic relaxation rate and reduce the
amplitude of MO caused by the inelastic mechanism.

Concluding, for observation of MO of $\sigma _{xx}$ caused by both of the
mechanisms (displacement and inelastic) in the 2D electron system formed on
the free surface of liquid helium, it is necessary to use low electron
densities and the parameters of the MW field giving a largest value
of $\lambda _{mw}$ defined by Eq.~(\ref{e15}).

In summary, we have investigated the influence of strong Coulomb forces
acting between SEs in liquid helium on photon-assisted scattering and on the
displacement mechanism of MW-induced oscillations of magnetoconductivity.
Coulomb interaction is shown to suppress strongly the amplitude of
oscillations and affect their shape. Under conditions of the experiment
with SE on liquid helium~\cite{KonKon-2010}, the amplitude of MO caused by
photon-assisted scattering is shown to be very small, which explains why
oscillations were not detected for MW frequencies
substantially different from the inter-subband resonance frequency. The
relationship between the amplitude of MO and parameters characterized the MW
field and the electron system obtained here allows to formulate conditions
under which photon-assisted scattering could be observed. In order to obtain
the same effect as in heterostructures, one need to increase the MW field
amplitude by a factor of about $3.9$. Therefore, the system of SEs on
liquid helium can serve as a model system for testing mechanisms of MO and
ZRS proposed for a 2D electron gas in heterostructures.

The work was partially supported by a Grant of SFFR and JSPS (F52.2/005).


\begin{thebibliography}{9}

\bibitem{ZudSim-2001} M.A. Zudov, R.R. Du, J.A. Simmons, and J.R. Reno,
Phys. Rev. B {\textbf{64}}, 201311(R) (2001).

\bibitem{ManSme-2002} R. Mani, J.H. Smet, K. von Klitzing, V. Narayanamurti,
W.B. Johnson, and V. Umansky, Nature \textbf{420}, 646 (2002).

\bibitem{ZudDu-2003} M.A. Zudov, R.R. Du, L.N. Pfeiffer, and K.W. West,
Phys. Rev. Lett. \textbf{90}, 046807 (2003).

\bibitem{DurSac-2003} A.C. Durst, S. Sachdev, N. Read, and S.M. Girvin, Phys. Rev.
Lett. \textbf{91}, 086803 (2003).

\bibitem{RyzSur-2003} V. Ryzhii and R. Suris, J. Phys.: Cond. Matt. \textbf{15}, 6855
(2003).

\bibitem{Shi-2003} V. Shikin, \textit{Pis'ma Zh. Eksp. Teor Fiz.} \textbf{77}, 281 (2003)
[JETP Lett. \textbf{77}, 236 (2003)].

\bibitem{KouRai-2003} A. A. Koulakov and M. E. Raikh, Phys. Rev. B \textbf{68},
115324 (2003).

\bibitem{RyzChaSur-2004} V. Ryzhii, A. Chaplik, R. Suris, \textit{Pis'ma Zh. Eksp. Teor Fiz.}
\textbf{80}, 412 (2004) [JETP Lett. \textbf{80}, 363 (2004)].

\bibitem{DmiVav-2005} I.A. Dmitriev, M.G. Vavilov, I.L. Aleiner, A.D. Mirlin, and
D.G. Polyakov, Phys. Rev. B \textbf{71}, 115316 (2005).

\bibitem{InaPla-2007} J. Inarrea and G. Platero, Phys. Rev. B \textbf{76}, 073311
(2007).

\bibitem{Mik-2011} S.A. Mikhailov, Phys. Rev. B \textbf{83}, 155303 (2011).

\bibitem{AndAle-2003} A.V. Andreev, I.L. Aleiner, and A.J. Millis, Phys.
Rev. Lett., \textbf{91}, 056803 (2003).

\bibitem{KonKon-2009} D. Konstantinov and K. Kono, Phys. Rev. Lett. \textbf{103},
266808 (2009).

\bibitem{KonKon-2010} D. Konstantinov and K. Kono, Phys. Rev. Lett. \textbf{105},
226801 (2010).

\bibitem{Mon-2011} Yu.P. Monarkha, \textit{Fiz. Nizk. Temp.} \textbf{37}, 108 (2011)
[Low Temp. Phys. \textbf{37}, 90 (2011)]; Yu.P. Monarkha,
\textit{Fiz. Nizk. Temp.} \textbf{37}, 829 (2011) [Low Temp. Phys. \textbf{37}, 655 (2011)].

\bibitem{Mon-2012} Yu.P. Monarkha, \textit{Fiz. Nizk. Temp.} \textbf{38}, 579 (2012)
[Low Temp. Phys. \textbf{38}, 451 (2012)].

\bibitem{KonMonKon-2013} D. Konstantinov, Yu.P. Monarkha, and K. Kono, Phys. Rev. Lett.
\textbf{111}, 266802 (2013).

\bibitem{Ryz-1969} V. I. Ryzhii, \textit{Fiz. Tverd. Tela} \textbf{11}, 2577 (1969)
[Sov. Phys. Solid State \textbf{11}, 2078 (1970)];

\bibitem{Ryz-2003} V. Ryzhii, Phys. Rev. B \textbf{68}, 193402 (2003).

\bibitem{Dor-2003} S. I. Dorozhkin, \textit{Pis'ma Zh. Eksp. Teor. Fiz.} \textbf{77}, 681
(2003) [JETP Lett. \textbf{77}, 577 (2003)].

\bibitem{BasLev-1965} F.G. Bass and I.B. Levinson, \textit{Zh. Eksp. Teor. Fiz.} \textbf{49}, 914 (1965).

\bibitem{MonKon-book} Yu.P. Monarkha and K. Kono, \textit{Two-Dimensional
Coulomb Liquids and Solids}, Springer-Verlag, Berlin Heildelberg (2004).

\bibitem{MonTesWyd-2002} Yu.P. Monarkha, E. Teske, and P. Wyder,
Phys. Rep. \textbf{370}, No. 1, pp. 1-61 (2002).

\bibitem{Mon-2013} Yu.P. Monarkha, \textit{Fiz. Nizk. Temp.} \textbf{39}, 1068 (2013)
[Low Temp. Phys. \textbf{39}, 828 (2013)].

\bibitem{Ger-1976} R.R. Gerhardts, Surf. Sci. \textbf{58}, 227 (1976).

\bibitem{AndUem-1974} T. Ando and Y. Uemura, J. Phys. Soc. Jpn. \textbf{36}, 959 (1974).

\bibitem{DykKha-1979} M.I. Dykman and L.S. Khazan, \textit{Zh. Eksp. Teor. Fiz.}
\textbf{77}, 1488 (1979) [Sov. Phys. JETP \textbf{50}, 747 (1979)].

\bibitem{FanDykLea-1997}  C. Fang-Yen, M.I. Dykman, and M.J. Lea, Phys. Rev. B
\textbf{55}, 16272 (1997).

\bibitem{RivSch-2004} P.H. Rivera and P.A. Schulz,  Phys. Rev. B \textbf{70}, 075314 (2004).

\end{thebibliography}
\end{document}